\documentclass{article} % For LaTeX2e
\usepackage{iclr2020_conference,times}

\usepackage{amsmath,amsfonts,bm}

\def\eqref#1{equation~\ref{#1}}
\def\1{\bm{1}}

\DeclareMathAlphabet{\mathsfit}{\encodingdefault}{\sfdefault}{m}{sl}
\SetMathAlphabet{\mathsfit}{bold}{\encodingdefault}{\sfdefault}{bx}{n}

\usepackage[utf8]{inputenc} % allow utf-8 input
\usepackage[T1]{fontenc}    % use 8-bit T1 fonts
\usepackage{hyperref}       % hyperlinks
\usepackage{url}            % simple URL typesetting
\usepackage{booktabs}       % professional-quality tables
\usepackage{amsfonts}       % blackboard math symbols
\usepackage{nicefrac}       % compact symbols for 1/2, etc.
\usepackage{microtype}      % microtypography
\usepackage{graphicx}
\usepackage{caption}
\usepackage{subcaption}
\usepackage{amsmath}
\usepackage{listings}
\usepackage{color}
\usepackage{enumitem}
\usepackage{algorithm2e}

\definecolor{codegreen}{rgb}{0,0.6,0}
\definecolor{codegray}{rgb}{0.5,0.5,0.5}
\definecolor{codepurple}{rgb}{0.58,0,0.82}
\definecolor{backcolour}{rgb}{0.95,0.95,0.92}

\lstdefinestyle{mystyle}{
    backgroundcolor=\color{backcolour},
    commentstyle=\color{codegreen},
    keywordstyle=\color{magenta},
    numberstyle=\tiny\color{codegray},
    stringstyle=\color{codepurple},
    basicstyle=\footnotesize,
    breakatwhitespace=false,
    breaklines=true,
    captionpos=b,
    keepspaces=true,
    numbers=left,
    numbersep=5pt,
    showspaces=false,
    showstringspaces=false,
    showtabs=false,
    tabsize=2,
    aboveskip=0em,
    belowcaptionskip=0em,
    belowskip=0em,
}

\newcommand{\property}{property\xspace}
\newcommand{\propertys}{properties\xspace}

\newcommand{\Propertys}{Properties\xspace}
\newcommand{\propertysig}{property signature\xspace}
\newcommand{\propertysigs}{property signatures\xspace}

\newcommand{\Propertysigs}{Property signatures\xspace}

\newcommand{\PropertySigs}{Property Signatures\xspace}

\newcommand{\True}{\ensuremath{\mathsf{True}}}
\newcommand{\False}{\ensuremath{\mathsf{False}}}

\newcommand{\alltrue}{\ensuremath{\mathsf{AllTrue}}}
\newcommand{\allfalse}{\ensuremath{\mathsf{AllFalse}}}
\newcommand{\mixed}{\ensuremath{\mathsf{Mixed}}}

\newcommand{\sig}{\mathrm{sig}}

\newcommand{\todo}[1]{}

\title{Learning to Represent Programs \\ with \PropertySigs}

\author{Augustus Odena, Charles Sutton \\
Google Research\\
\texttt{\{augustusodena,charlessutton\}@google.com}
}

\iclrfinalcopy % Uncomment for camera-ready version, but NOT for submission.
\begin{document}

\maketitle

\begin{abstract}
We introduce the notion of \propertysigs, a representation for programs
and program specifications meant for consumption by machine learning algorithms.
Given a function with input type $\tau_{in}$ and output type
$\tau_{out}$, a \property is a function of type:
$(\tau_{in}, \tau_{out}) \rightarrow \texttt{Bool}$ that (informally) describes
some simple property of the function under consideration.
For instance, if $\tau_{in}$ and $\tau_{out}$ are both lists of the same type,
one \property might ask `is the input list the same length as the output list?'.
If we have a list of such \propertys, we can evaluate them all for our
function to get a list of outputs that we will call the  \propertysig.
Crucially, we can `guess' the \propertysig for a function given only
a set of input/output pairs meant to specify that function.
We discuss several potential applications of \propertysigs and show
experimentally that they can be used to improve over a baseline synthesizer
so that it emits twice as many programs in less than one-tenth of the time.
\end{abstract}

\section{Introduction}

Program synthesis is a longstanding goal of computer science research
\citep{TOWARDAUTOMATIC, PROW, LISP, LISP2, REACTIVE, KNOWLEDGEANDREASONING},
arguably dating to the 1940s and 50s \citep{TURING, FORTRAN}.
Deep learning methods have shown promise at
automatically generating programs from a small
set of input-output examples
\citep{DEEPCODER,devlin17a,ellis18library,ellis19}.
In order to deliver on this promise, we believe it is important to represent
programs and specifications in a way that supports learning.
Just as computer vision methods benefit from
the inductive bias inherent to convolutional neural networks \citep{CONVNETS},
and likewise with LSTMs for natural language and other sequence data \citep{LSTM},
it stands to reason that ML techniques for computer programs will benefit
from architectures with a suitable inductive bias.

We introduce a new representation for programs and their
specifications, based on the principle that
\emph{to represent a program, we can use a set of simpler programs}.
This leads us to introduce the concept of a  \property,
which is a program that computes a boolean function of the
input and output of another program.
For example, consider the problem of synthesizing
a program from a small set of input-output examples.
Perhaps the synthesizer is given a few pairs of lists of integers,
and the user hopes that the synthesizer will produce a sorting function.
Then useful \propertys might include functions
that check if the input and output lists have the same length,
if the input list is a subset of the output, if element $0$ of the output
list is less than element $42$, and so on.

The outputs of a set of \propertys can be concatenated into a
vector, yielding a representation that we call a \emph{\propertysig}.
\Propertysigs can then be used for consumption by machine learning algorithms,
essentially serving as the first layer of a neural network.
In this paper, we demonstrate the utility of \propertysigs for program synthesis,
using them to perform a type of premise selection as in \citet{DEEPCODER}.
More broadly, however, we envision that \propertysigs could be useful across a broad range of problems,
including algorithm induction \citep{devlin17a},
improving code readability \citep{naturalize}, and program analysis \citep{heo19continuously}.

More specifically, our contributions are:

\begin{itemize}[wide]
\item We introduce the notion of \propertysigs, which are
  a general purpose way of featurizing both programs and program specifications
  (Section~\ref{sec:properties}).

\item We demonstrate how to use \propertysigs within a machine-learning based synthesizer
  for a general-purpose programming language. This allows us to automatically learn
  a useful set of \propertysigs, rather than choosing them manually (Sections~\ref{sec:learning-properties}
  and~\ref{sec:synthesis-results}).

  \item We show that a machine learning model can predict the signatures of individual
  functions given the signature of their composition, and describe several ways this
  could be used to improve existing synthesizers (Section~\ref{sec:composition}).

\item We perform experiments on a new test set of 185 functional
  programs of varying difficulty, designed to be the sort of algorithmic
  problems that one would ask on an undergraduate computer science examination.
  We find that the use of \propertysigs leads to a dramatic
  improvement in the performance of the synthesizer, allowing it to synthesize
  over \emph{twice as many programs in less than one-tenth of the time}
  (Section~\ref{sec:synthesis-results}).
    An example of a complex program that was synthesized only by the \propertysigs
    method is shown in Listing~\ref{listing:introexample}.
\end{itemize}

For our experiments, we created a specialized programming language,
called Searcho\footnote{
Searcho is heavily based on code written by Niklas Een, which is
available at
\url{https://github.com/tensorflow/deepmath/tree/master/deepmath/zz/CodeBreeder}
}
(Section~\ref{sec:searcho}),
based on strongly-typed functional languages such as Standard ML and Haskell.
Searcho
is designed so that many similar programs can be executed
rapidly, as is needed during a large-scale
distributed search during synthesis.
We release\footnote{
Available at \url{https://github.com/brain-research/searcho}
} the programming language, runtime environment,
distributed
search infrastructure, machine learning models, and training data from our
experiments so that they can be used for future research.

\begin{lstlisting}[language=Caml, caption={
A program synthesized by our system, reformatted and with
variables renamed for readability.
This program returns the sub-list of all of the elements in a list
that are distinct from their previous value in the list.
}, label=listing:introexample]
fun unique_justseen(xs :List<Int>) -> List<Int> {
  let triple = list_foldl_<Int, (List<Int>, Int, Bool)>(
    xs,
    (nil<Int>, 0, _1),
    \(list_elt, (acc, last_elt, first)){
      cond_(or_(first, not_equal_(list_elt, last_elt)),
        \{(cons_(list_elt, acc), list_elt, _0)},
        \{(acc                 , list_elt, _0)})
    });
  list_reverse_(#0(triple))
};
\end{lstlisting}

\section{Programming By Example and The Searcho Language}

In Inductive Program Synthesis, we are given a specification of a program
and our goal is to synthesize a program meeting that specification.
Inductive Synthesis is generally divided into Programming by Example (PBE) and
Programming by Demonstration (PBD).
This work is focused on PBE.
%% CS: Commenting out for space.
%% but we will describe both for context.
%% In PBD, we are given some
%% form of execution trace for the desired program.
%% This may be a sequence of primitive function calls or a stack trace or something
%% else. Examples of PBD systems include \citet{PYGMALION}, \citet{LISP}, and
%% \citet{PBD}.
In PBE, we are given a set of input/output pairs such that for each pair,
the target program takes the input to the corresponding output.
Existing PBE systems include \citet{PBE}, \citet{MLPBE}, and \citet{FLASHFILL}.
A PBE specification might look like
Listing~\ref{listing:squarespec}:

\begin{lstlisting}[language=Haskell, caption={
An example PBE specification.
}, label=listing:squarespec]
io_pairs = [(1, 1), (2, 4), (6, 36), (10, 100)]
\end{lstlisting}

for which a satisfying solution would be the function squaring its input.
Arbitrarily many functions satisfy this specification.
It is interesting but out of scope\footnote{
Though note that in this work and in prior work, the search procedure
used will tend to emit `shorter' programs first, and so there is an
Occam's-Razor-type argument \citep{OCCAM} to be made that you \textit{should}
get this for free.}
to think about ways to ensure that the synthesis procedure recovers
the `best' or `simplest' program satisfying the specification.

\label{sec:searcho}
Much (though not all) work on program synthesis is focused
on domain specific languages that are less than maximally expressive
\citep{FLASHFILL, DEEPCODER, SQL, STUN}.
We would like to focus on the synthesis of programs in a Turing complete language, but
this presents technical challenges:
First, general purpose languages such as C++ or Python are typically quite complicated
and sometimes not fully specified; this makes it a
challenge to search over partial programs in those languages.
Second, sandboxing and executing code written in these languages is nontrivial.
Finally, searching over and executing many programs in these languages can be
quite slow, since this is not what they were designed for.

For these reasons, we have created a general-pupose, Turing complete
programming language and runtime.
The programming language is called Searcho and it and its runtime have been
designed specifically with program synthesis in mind.
The language can roughly be thought of as a more complicated version of the
simply typed lambda calculus or as a less complicated version of Standard ML or
OCaml.\footnote{
In this paper, we will present illustrative programs in Haskell syntax
  to make them more broadly readable.
  Searcho programs will be presented in Searcho syntax, which is similar.
  }
Searcho code is compiled to bytecode and run on the Searcho Virtual Machine.
Code is incrementally compiled, which means that the standard library and
specification can be compiled once and then many programs can be pushed on and
popped off from the stack in order to check them against the specification.
Searcho is strongly typed with algebraic datatypes \citep{TAPL}\footnote{
Types have been shown to
substantially speed up synthesis. See e.g. Figure 6 of \citet{LAMBDASQUARED}.}
Searcho includes a library of 86 functions, all of which are supported by our
synthesizer. This is a significantly larger
language and library than have been used in previous work on neural program synthesis.

We have also implemented a baseline enumerative synthesizer.
The main experiments in this paper will involve plugging the outputs
of a machine learning model into the configuration for our baseline synthesizer
to improve its performance on a set of human-constructed PBE tasks.

\section{\PropertySigs}
\label{sec:properties}

Consider the PBE specification in Listing~\ref{listing:concatwithreverse}:

\begin{lstlisting}[language=Haskell, caption={
An example PBE Specification.
}, label=listing:concatwithreverse]
io_pairs = [
  ([1, 2345, 34567],   [1, 2345, 34567, 34567, 2345, 1]),
  ([True, False],      [True, False, False, True]),
  (["Batman"],         ["Batman", "Batman"]),
  ([[1,2,3], [4,5,6]], [[1,2,3], [4,5,6], [4,5,6], [1,2,3]])
]
\end{lstlisting}

We can see that the function concatenating the input list to its reverse
will satisfy the specification, but how can we teach this to a computer?
Following \citet{DEEPCODER} we take the approach of training a machine learning
model to do premise selection for a symbolic search procedure.
But how do we get a representation of the specification to feed to the model?
In \citet{DEEPCODER}, the model acts only on integers and lists of integers,
constrains all integers to lie in $[-256, 256]$, has special-case handling of
lists, and does not deal with polymorphic functions.
It would be hard to apply this technique to the above specification,
since the first example contains unbounded integers, the second example contains
a different type than the first\footnote{
So any function satisfying the spec will be parametrically polymorphic.},
and the third and fourth examples
contain recursive data structures (lists of characters and lists of
integers respectively).

Thankfully, we can instead learn a representation that is composed of
the outputs of multiple other programs running on each input/output pair.
We will call these other programs \propertys.
Consider the three \propertys in Listing~\ref{listing:projections}.

\begin{lstlisting}[language=Haskell, caption={
Three function projections that can act on the specification from
Listing~\ref{listing:concatwithreverse}.
}, label=listing:projections]
all_inputs_in_outputs ins outs    = all (map (\x -> x in outs) ins)
ouputs_has_dups ins outs          = has_duplicates (outs)
input_same_len_as_output ins outs = (len ins) == (len outs)
\end{lstlisting}

Each of these three programs can be run on all 4 of the input output pairs to yield a \texttt{Boolean}.
The first always returns \True\ for our spec, as does the second.
The third always returns \False\ on the given examples, although note that
it would return \True\ if the examples had contained  the implicit base case of the empty list.
Thus, we can write that our spec has the `\propertysig' $[\True, \True, \False]$.

How is this useful?
From the first \property we can infer that we should not throw away any elements of the input list.
From the third we might guess that we have to add or remove elements from the input list.
Finally, the second might imply that we need to create copies of the input elements somehow.
This does not narrow our search down all the way, but it narrows it down quite a lot.
Since the \propertys are expressed in the same language as the programs we are synthesizing,
we can emit them using the same synthesizer.
Later on, we will describe how we enumerate many random \propertys and prune them to keep only the useful
ones.
The \propertysigs that we consider in our experiments contain thousands of values.

Since the output of these \propertys is either always True, always False, or sometimes True and
sometimes False, a neural network can learn embeddings for those three
values and it can be fed a vector of such values, one for each applicable
\property, as the representation of a program specification.

\subsection{Abstracting \Propertys into Signatures}

Now we describe our representation for a program $f :: \tau_{in} \rightarrow \tau_{out}$.
Each \property is a program $p :: (\tau_{in}, \tau_{out}) \rightarrow \texttt{Bool}$
that represents a single ``feature'' of the program's inputs and outputs
which might be useful for its representation.\footnote{Although we write $f$ as
a function, that is, as returning an output, it is easy to handle procedures
that do not return a value by defining $\tau_{out}$ to be a special void type.}
In this section, we assume that we have determined a sequence $P = [p_1 \ldots p_n]$ of \propertys
that are useful for describing $f$, and we wish to combine them into a single
representation of $f$. Later, we will describe a learning principle
for choosing relevant \propertys.

We want the \propertysig to summarize the output of all the properties in $P$
over all valid inputs to $f$. To do this, we first extend the notion of property
to a set of inputs in the natural way.
If $S$ is a set of values of type $\tau_{in}$ and $p \in P$, we define
$p(S) = \{ p(x, f(x)) \,|\, x \in S \}.$
Because $p(S)$ is a  set of booleans, it can have only three possible values, either
$p(S) = \{ \True \},$ or
$p(S) = \{ \False \},$ or
$p(S) = \{ \True, \False \}$, corresponding respectively to the cases
where $p$ is always true, always false, or neither.
To simplify notation slightly, we define the function $\Pi$
as
$\Pi(\{\True\}) = \alltrue$, $\Pi(\{\False\}) = \allfalse$, and $\Pi(\{\True, \False\}) = \mixed.$
Finally,
 we can define the \emph{\propertysig} $\sig(P, f)$ for a program $f$ and a \property sequence  $P$ 
as
\begin{align*}
  \sig(P, f)[i] &= \Pi(p_i(V(\tau_{in}))),
\end{align*}
where $V(\tau_{in})$ is the possibly infinite set of all values of type $\tau_{in}$.

Computing the property signature for $f$ could be intractable or undecidable,
as it might require proving difficult facts about the program.
Instead, in practice, we will compute an \emph{estimated \propertysig} for a small
set of input-output pairs $S_{io}.$
The estimated \propertysig summarizes the actions of $P$ on $S_{io}$ rather than
on the full set of inputs $V(\tau_{in})$.
Formally, the estimated \propertysig  is
\begin{equation}
  \widehat{\sig}(P, S_{io})[i] := \Pi(\{p_i(x_{in}, x_{out}) \,|\, (x_{in}, x_{out}) \in S_{io}\}).
\end{equation}
This estimate gives us an under-approximation of the true signature of $f$ in the following sense:
If we have $\widehat{\sig}(P, S) = \mixed$, we must also have $\sig(P, f) = \mixed$.
If $\widehat{\sig}(P, S) = \alltrue$, then either $\sig(P, f) = \alltrue$ or $\sig(P, f) = \mixed$,
and similarly with $\allfalse$.
Estimated \propertysigs are particularly useful for synthesis using PBE,
because we can compute them from the input-output pairs that specify
the synthesis task, without having the definition of $f$.
Thus we can use estimated \propertysigs to `featurize' PBE specifications
for use in synthesis.

\subsection{Learning Useful \Propertys}
\label{sec:learning-properties}

How do we choose a set of properties that will be useful for synthesis?
Given a training set of random programs with random input/output examples,
we generate many random \propertys.
We then prune the random \propertys based on whether they distinguish between
any of the programs.
Then, given a test suite of programs, we do an additional pruning
step: among all \propertys that give the same value for every element of the
test suite, we keep the shortest \property,
because of Occam's razor  considerations.
Given these `useful' properties, we can train a premise selector \citep{DEEPCODER} to predict
library function usage given properties. Specifically, from the remaining properties,
we compute estimated property signatures for each function in the training set,
based on its input output examples. Then we use the property signature as the
input to a feedforward network that predicts the number of times each library
function appears in the program.
In Section~\ref{sec:synthesis-results}, we will give more details about the architecture
of this premise selector, and evaluate it for synthesis.
For now, we point out that this premise selector could itself be used to
find useful properties, by examining which properties
are most useful for the model's predictions.

\subsection{Why are \PropertySigs Useful?}
Experiments in the next section will establish that \propertysigs
let our baseline synthesizer emit programs it previously could not, but we
think that they can have broader utility:

\begin{itemize}[wide]
\item They allow us to represent more types of functions.
\Propertysigs can automatically deal with unbounded data types,
recursive data types, and polymorphic functions.

\item They reduce dependency on the distribution from which examples
are drawn. If the user of a synthesizer gives example inputs distributed
differently than the training data, the `estimated' \propertys might not change
much.\footnote{
This argument does rely on \propertys being somehow simple.
For instance, if the \property does not compute whether a list contains the
value 777, it cannot fail to generalize with respect to the presence or absence
of 777.
Since we search for \propertys in a shortest-first fashion, the \propertys
we find should be biased toward simplicity, though certainly this hypothesis
merits more experimental validation.
}

\item They can be used wherever we want to search for functions by semantics.
Imagine a search engine where users give a specification, the
system guesses a \propertysig, and this signature guess is used to find
all the pre-computed functions with similar semantics.

\item Synthesized programs can themselves become new \propertys.
For example, once I learn a program for primality checking,
I can use primality checking in my library of \propertys.
\end{itemize}

\section{Program Synthesis with \propertysigs}
\label{sec:synthesis-results}

We design an experiment to answer the following question:
Can \propertysigs help us synthesize programs that we otherwise could not have synthesized?
As we will show, the answer is yes!

\subsection{Experimental Setup}

\paragraph{How Does the Baseline Synthesizer Work?}

Our baseline synthesizer is very similar to that in \citet{LAMBDASQUARED} and
works by filling in typed holes\footnote{
In the synthesis literature, this approach of first discovering the high-level
structure and then filling it in is sometimes called `top-down'
synthesis \citep{ARMANDOCOURSE}.
Top-down synthesis is to be contrasted with `bottom-up' synthesis, in
which low-level components are incrementally combined into larger programs.
}.
That is, we infer a program type $\tau_{in} \rightarrow \tau_{out}$ from the
specification and the synthesizer starts with a empty `hole' of type
$\tau_{in} \rightarrow \tau_{out}$ and then fills
it in all possible ways allowed by the type system.
Many of these ways of filling-in will yield new holes, which can in turn
be filled by the same technique.
When a program has no holes, we check if it satisfies the spec.
We order the programs to expand by their cost, where the cost is essentially a
sum of the costs of the individual operations used in the program.

At the beginning of the procedure, the synthesizer is given a configuration,
which is essentially a weighted set of \emph{pool elements} that it is allowed to use to fill in the holes.
A pool element is a rewrite rule that replaces a hole with a type-correct Searcho program,
which may itself contain its own, new holes.
In our synthesizer,
there is one possible pool element for each of the 86 library functions in Searcho,
which calls the library function, with correctly-typed holes for each of its arguments.
The configuration will specify a small subset of these pool elements to use during search.
It is through the configuration  that we will use machine learning to inform
the search procedure, as we describe later.
See Appendix \ref{app:baseline} for further details on this baseline system.

\paragraph{How is the Training Data Generated?}
Our test corpus contains programs with 14 different types.
For each of those 14 types, we randomly sample configurations
and then randomly generate training programs for each
configuration, pruning for observational equivalence.
We generate up 10,000 semantically distinct programs for each type, though of
course some function types admit less distinct programs than this (
e.g. $\texttt{Bool} \rightarrow \texttt{Bool}$).
We also generate and prune random properties as described in
Section \ref{sec:learning-properties}.
See Listing \ref{listing:usefulproperties} for examples of useful \propertys
that were generated.

\begin{lstlisting}[language=Haskell, caption={
4 of the Properties with the highest discriminative power on functions of type
$\texttt{List<Int>} \rightarrow \texttt{List<Int>}$.
The first checks if every element of the input list is in the output list.
The second checks if the length of the output list is even.
The third checks if sum of the input and the output list is the same, and the
fourth checks if the input list is longer than the output list.
}, label=listing:usefulproperties]
\:(List<Int>, List<Int>)->Bool (input, output) {
  list_for_all_<Int> (input, \x {in_list_<Int> (x, output)})}
\:(List<Int>, List<Int>)->Bool (input, output) {
  not_ (is_even_ (list_len_<Int> output))}
\:(List<Int>, List<Int>)->Bool (input, output) {
  not_equal_<Int> ((ints_sum_ input), (ints_sum_ output))}
\:(List<Int>, List<Int>)->Bool (input, output) {
  gt_ ((list_len_<Int> input), (list_len_<Int> output))}
\end{lstlisting}

\paragraph{How was the Test Set Constructed?}
We've constructed a test set of 185 human generated programs
ranging in complexity from one single line to many nested function
calls with recursion.
Programs in the test set include computing the GCD of two integers, computing
the $n$-th fibonacci number, computing the intersection of two sets, and
computing the sum of all pairs in two lists.
We ensure that none of the test functions appear in the training set.
See the open source code for more details on this.

\paragraph{What is the Architecture of the Model?}
As mentioned above, we train a neural network to predict the number of times
each pool element will appear in the output.
This neural network is fully connected, with learned embeddings for each of the
values $\alltrue$, $\allfalse$ and $\mixed$.

\paragraph{How does the Model Output Inform the Search Procedure?}
Since we have a large number of pool elements (86), we can't run the synthesizer
with all pool elements if we want to find programs of reasonable length.
This is both because we will run out of memory and because it will take too long.
Thus, we randomly sample configurations with less pool elements.
We then send multiple such configurations to a distributed synthesis server that
tries them in parallel.

When we use the model predictions, we sample pool elements in proportion
to the model's predicted number of times that pool element appears.
The baseline samples pool elements in proportion to their rate of appearance in
the training set.

\subsection{Using \PropertySigs Lets us Synthesize New Functions}
\begin{figure}[tb]
  \centering
  \includegraphics[width=0.7\textwidth]{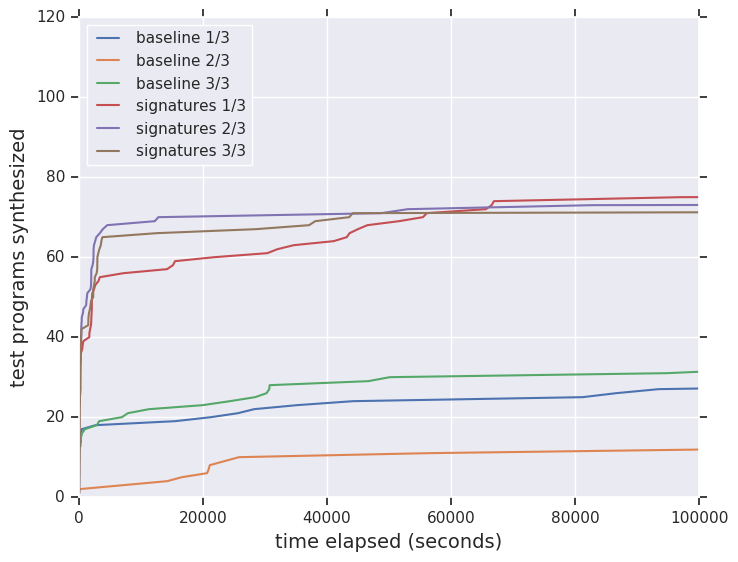}
  \caption{
    Comparison of synthesis with \propertysigs and without \propertysigs.
    The $x$-axis denotes time elapsed in seconds.
  Roughly speaking, we let the distributed synthesizer run for 1 day.
  The $y$-axis represenets the cumulative number of programs synthesized.
  On average, the baseline solved 28 of the test programs, while the baseline
  enhanced with \propertysigs solved 73 test programs (around 2.6 times
  as many programs).
  Both the baseline and the run with \propertysigs were run with three
  different random seeds.
  Altogether, this experiment provides strong evidence that
  \propertysigs can be useful.
  }\label{fig:main}
\label{fig:svhn}
\end{figure}

We ran 3 different runs of our distributed synthesizer for 100,000 seconds
with and without the aid of \propertysigs.
The baseline synthesizer solved 28 test programs on average.
With \propertysigs, the synthesizer solved an average of 73 test
programs.
See Figure \ref{fig:main} for more discussion.
Indeed, it can be seen from the figure that not only did the synthesizer
solve many more test programs using \propertysigs, but it did so much faster,
synthesizing over twice as many programs in one-tenth of the time as the
baseline.

\subsection{Comparison with DeepCoder}

We have conducted an experiment to compare premise selection using Property
Signatures to the premise selection algorithm from \citep{DEEPCODER}.
This required considerable modifications to the experimental procedure.

First, since the premise-selection part of DeepCoder can only handle Integers
and lists of Integers, we restricted the types of our training \textbf{and} test
functions. In particular, we read through \citep{DEEPCODER} and found four
function types in use:

\begin{lstlisting}[language=Haskell, caption={
The four function types used in DeepCoder.
}, label=listing:types]
f :: [Int] -> [Int]
g :: [Int] -> Int
h :: ([Int], [Int]) -> Int
k :: ([Int], Int) -> Int
\end{lstlisting}

The types of f and g in \ref{listing:types} are taken directly from
\citep{DEEPCODER}. The types of h and k are inferred from examples given in the
appendix of \citep{DEEPCODER}. Their DSL does not technically have tuples,
but we have wrapped the inputs of their `two-input-functions' in tuples
for convienence.

Second, since DeepCoder can only handle integers betwen $-255$ and $255$, we
first re-generated all of our random inputs (used for `hashing' of generated
training data) to lie in that range. We then generated random training functions
of the above four types. We then made a data set of training functions
associated with 5 input-output pairs,
throwing out pairs where any of the outputs
were outside the aforementioned range, and throwing out functions where
\textit{all} outputs contained some number outside that range.

Third, of the examples in our test set with the right types, we modified their
input output pairs in a similar way.
We filtered out functions that could not be so modified.
After doing so, we were left with a remaining test suite of 32 functions.

Finally, we trained a model to predict functions-to-use from learned embeddings
of the input-output pairs, as in DeepCoder.
We didn't see a description of how functions with multiple inputs had their
inputs embedded, so we elected to separate them with a special character,
distinct from the null characters that are used to pad lists.

Compared with the Property Signatures method, this technique results in far
fewer synthesized test set programs.
We did 3 random restarts for each of DeepCoder, Property Signatures, and the
Random Baseline (recall that the random baseline itself is already a relatively
sophisticated synthesis algorithm - it's just the configurations that are
random).
The 3 DeepCoder runs synthesized an average
of $3.33$ test programs, while the Property Signature runs (trained on the same
modified training data and tested on the same modified test data) synthesized
$16.33$. The random baseline synthesized $3$ programs on average.

A priori, this seems like a surprisingly large gap, but it actually fits with
what we know from existing literature. \citet{SHIFT} observe something similar:
which is that DeepCoder-esque techniques tend to generalize poorly to a
a test set where the input-output pairs come from a different distribution
than they do in training.
This is the case in our experiment, and it will be the case in any realistic
setting, since the test set will be provided by users.
Property Signatures are (according to our experiments) much less sensitive to
such shift.
This makes intuitive sense: whether an input list is half the length of an
output list (for instance) is
invariant to the particular distribution of members of the list.

Note that even if Property Signatures did not outperform DeepCoder on this
subset of our test set, they would still constitute an improvement due to their
allowing us to operate on arbitrary programs and inputs types.

\section{Predicting \propertysigs of Function Compositions}
\label{sec:composition}

Most programs involve composing functions with other functions.
Suppose that we are trying to solve a synthesis problem from
a set of input/output examples, and during the search
we create a partial program of the form $f (g (x))$ for
some unknown $g$.
Since we know $f$, we know its \propertysig.
Since we have the program specification, we also have the estimated \propertysig
for $f \circ g := f(g(x))$.
If we could somehow guess the signature for $g$, we could look it up in a
cache of previously computed functions keyed by signature.
If we found a function matching the desired signature, we would be done.
If no matching function exists in the cache, we could start a smaller search with
only the signature of $g$ as the target, then use that result in our original
search.
We could attempt to encode the relationship between $f$ and $g$ into a set of
formal constraints and pass that to a solver of some kind \citep{Z3},
and while that is potentially an effective approach, it may be difficult
to scale to a language like Searcho.
Instead, we can simply train a machine learning model to predict the signature
of $g$ from the signature of $f$ and the signature of $ f \circ g$.

Here we present an experiment to establish a proof of concept of this idea.
First, we generated a data set of 10,000 random functions taking lists of
integers to lists of integers.
Then we randomly chose 50,000 pairs of functions from this list, arbitrarily
designating one as $f$ and one as $g$.
We then computed the signatures of $f$, $g$ and $f \circ g$ for each pair,
divided the data into a training set of 45,000 elements and a test set of
5,000 elements, and trained a
small fully connected neural network to predict the signature of $g$ from the
other two signatures.

On the test set, this model had 87.5\% accuracy, which is substantially better
than chance.
We inspected the predictions made on the test set and found
interesting examples like the one in Listing \ref{listing:composition}, where
the model has learned to do something you might (cautiously) refer to as logical
deduction on properties. This result is suggestive of the expressive power of
\propertysigs.
It also points toward exciting future directions for research into neurally
guided program synthesis.

\begin{lstlisting}[language=Haskell, caption={
Example of successful prediction made by our composition predictor model.
The property in question checks whether all the elements of the output list
are members of the input list. For $f$, the value is \alltrue, and for
$f \circ g$ the value is \mixed. The model doesn't know $g$ or its signature,
but correctly predicts that the value of this property for $g$ must be \mixed.
}, label=listing:composition]
f: \:List<Int>->List<Int> inputs {
  consume_ (inputs, (list_foldl_<Int, Int> (inputs, int_min, mod_)))}
g: \:List<Int>->List<Int> inputs {
  list_map_<Int, Int> (inputs, neg_)}
prop: \:(List<Int>, List<Int>)->Bool (inputs, outputs) {
  list_for_all_<Int> (outputs, \x {in_list_<Int> (x, inputs)})}
\end{lstlisting}

\section{Related Work}
There is substantial prior work on program synthesis in general.
We can hardly do it justice here, but see some of
\citet{THREEPILLARS,ARMANDOCOURSE,RISHABHSURVEY,BIGCODE} for more detailed
surveys.

\paragraph{Property Based Testing:}
Function \propertys are similar to the properties from Property Based Testing,
a software testing methodology popularized by the QuickCheck library
\citep{QUICKCHECK} that has now spread to many contexts
\citep{QCRUST,QCJAVA,QCPYTHON,AFLPLUSQUICKCHECK,PBTIF,WIPBT}.
Quickcheck properties are human-specified and operate on functions,
while our \propertys operate on input/output pairs.

\paragraph{Automated Theorem Proving:}
Synthesizing programs using machine learning is related to the idea of proving
theorems using machine learning \citep{DEEPMATH}.
Synthesis and theorem proving are
formally related as well \citep{CURRYHOWARD}.

\paragraph{Program Synthesis from a Programming Languages Perspective:}
Most existing work on synthesis approaches is from the perspective of
programming language design.
Our baseline synthesizer borrows many ideas from \cite{LAMBDASQUARED}.
\cite{SYNQUID} use refinement types \citep{REFINEMENTTYPES}
(roughly, a decidable version of dependent types - see \citet{TAPL}) to give
program specifications, allowing the type-checker to discard many candidate
programs.
\Propertysigs can be thought of as a compromise between refinement
types and dependent types: we can write down specifications with them that would
be impossible to express in refinement types, but we can only check those
specifications empirically.

\paragraph{ML-Guided Program Synthesis:}
More recently, researchers have used machine learning to synthesize and
understand programs.
We have mentioned \citet{DEEPCODER}, but see all of:
\citet{INFERSKETCHES,ECSQUARED,GARBAGECOLLECTOR,NGDS,WRITEEXECUTEASSESS,HIERARCHICALBAYES,CODE2VEC} as well.
\citet{MLPBE} introduces the idea of features: a predecessor
to the idea of \propertys.
Features differ from \propertys in that they are hand-crafted rather than
learned, and that they were applied only on a limited string processing domain.

\paragraph{Deepcoder:}
The relationship between this work and \citet{DEEPCODER} merits special
discussion.
Aside from the inclusion of \propertysigs,
they differ in the following ways:
\begin{itemize}[wide]
% \begin{itemize}[wide]

\item We use a more expressive DSL.
Their DSL only allows linear control flow with a small set of functions,
whereas our language is Turing complete (it has looping, recursion, etc).
We also have a larger set of allowed component functions: 86 vs. 34.

\item Their machine learning method does not work straightforwardly for
arbitrary programs.
Their training and test programs only deal with integers and lists of integers,
while we have 14 different function types.
It would thus not be feasible to compare the techniques on anything but a tiny subset
of our existing test set.

\item The test cases in \citet{DEEPCODER} are generated from their enumerative
synthesizer. It is therefore guaranteed that the synthesizer will be able to emit
them in a reasonable amount of time during testing, so their demonstrated
improvements are `merely' speed-ups.
Our test cases are human generated, and over half of the programs synthesized
using \propertysigs were not synthesized at all\footnote{
Of course, barring bugs in the synthesizer, they would be synthesized
\textit{eventually}.
} given over a day of time.

\end{itemize}

\section{Conclusion and Future Work}
In this work, we have introduced the idea of \propertys and
\propertysigs.
We have shown that \propertysigs allow us to synthesize programs that
a baseline otherwise was not able to synthesize, and have sketched out
other potential applications as well.
Finally, we have open sourced all of our code, which we hope will accelerate
future research into ML-guided program synthesis.

\subsubsection*{Acknowledgments}
We would like to thank Kensen Shi, David Bieber, and the rest of the Program
Synthesis Team for helpful discussions.
We would like to thank Colin Raffel for reading a draft of the paper.
Most of all, we owe a substantial debt to Niklas Een, on whose Evo programming
language (\url{https://github.com/tensorflow/deepmath/tree/master/deepmath/zz/CodeBreeder})
the Searcho language is heavily based.
\newpage
\newpage

\bibliography{iclr2020_conference}

\begin{thebibliography}{48}
\providecommand{\natexlab}[1]{#1}
\providecommand{\url}[1]{\texttt{#1}}
\expandafter\ifx\csname urlstyle\endcsname\relax
  \providecommand{\doi}[1]{doi: #1}\else
  \providecommand{\doi}{doi: \begingroup \urlstyle{rm}\Url}\fi

\bibitem[Allamanis et~al.(2014)Allamanis, Barr, Bird, and Sutton]{naturalize}
Miltiadis Allamanis, Earl~T Barr, Christian Bird, and Charles Sutton.
\newblock Learning natural coding conventions.
\newblock In \emph{Symposium on the Foundations of Software Engineering (FSE)},
  2014.

\bibitem[Allamanis et~al.(2018)Allamanis, Barr, Devanbu, and Sutton]{BIGCODE}
Miltiadis Allamanis, Earl~T Barr, Premkumar Devanbu, and Charles Sutton.
\newblock A survey of machine learning for big code and naturalness.
\newblock \emph{ACM Computing Surveys (CSUR)}, 51\penalty0 (4):\penalty0 81,
  2018.

\bibitem[Alon et~al.(2019)Alon, Zilberstein, Levy, and Yahav]{CODE2VEC}
Uri Alon, Meital Zilberstein, Omer Levy, and Eran Yahav.
\newblock code2vec: Learning distributed representations of code.
\newblock \emph{Proceedings of the ACM on Programming Languages}, 3\penalty0
  (POPL):\penalty0 40, 2019.

\bibitem[Alur et~al.(2015)Alur, {\v{C}}ern{\'y}, and Radhakrishna]{STUN}
Rajeev Alur, Pavol {\v{C}}ern{\'y}, and Arjun Radhakrishna.
\newblock Synthesis through unification.
\newblock In Daniel Kroening and Corina~S. P{\u{a}}s{\u{a}}reanu (eds.),
  \emph{Computer Aided Verification}, pp.\  163--179, Cham, 2015. Springer
  International Publishing.
\newblock ISBN 978-3-319-21668-3.
\newblock URL
  \url{http://ecee.colorado.edu/pavol/publications/cav15a/cav15a.pdf}.

\bibitem[Backus et~al.(1957)Backus, Beeber, Best, Goldberg, Haibt, Herrick,
  Nelson, Sayre, Sheridan, Stern, Ziller, Hughes, and Nutt]{FORTRAN}
J.~W. Backus, R.~J. Beeber, S.~Best, R.~Goldberg, L.~M. Haibt, H.~L. Herrick,
  R.~A. Nelson, D.~Sayre, P.~B. Sheridan, H.~Stern, I.~Ziller, R.~A. Hughes,
  and R.~Nutt.
\newblock The fortran automatic coding system.
\newblock In \emph{Papers Presented at the February 26-28, 1957, Western Joint
  Computer Conference: Techniques for Reliability}, IRE-AIEE-ACM '57 (Western),
  pp.\  188--198, New York, NY, USA, 1957. ACM.
\newblock \doi{10.1145/1455567.1455599}.
\newblock URL \url{http://doi.acm.org/10.1145/1455567.1455599}.

\bibitem[Balog et~al.(2016)Balog, Gaunt, Brockschmidt, Nowozin, and
  Tarlow]{DEEPCODER}
Matej Balog, Alexander~L Gaunt, Marc Brockschmidt, Sebastian Nowozin, and
  Daniel Tarlow.
\newblock Deepcoder: Learning to write programs.
\newblock \emph{arXiv preprint arXiv:1611.01989}, 2016.

\bibitem[Claessen \& Hughes(2011)Claessen and Hughes]{QUICKCHECK}
Koen Claessen and John Hughes.
\newblock Quickcheck: a lightweight tool for random testing of haskell
  programs.
\newblock \emph{Acm sigplan notices}, 46\penalty0 (4):\penalty0 53--64, 2011.

\bibitem[Copeland(2012)]{TURING}
B.J. Copeland.
\newblock \emph{Alan Turing's Electronic Brain: The Struggle to Build the ACE,
  the World's Fastest Computer}.
\newblock OUP Oxford, 2012.
\newblock ISBN 9780199609154.
\newblock URL \url{https://books.google.com/books?id=YhQZnczOS7kC}.

\bibitem[De~Moura \& Bj{\o}rner(2008)De~Moura and Bj{\o}rner]{Z3}
Leonardo De~Moura and Nikolaj Bj{\o}rner.
\newblock Z3: An efficient smt solver.
\newblock In \emph{International conference on Tools and Algorithms for the
  Construction and Analysis of Systems}, pp.\  337--340. Springer, 2008.

\bibitem[Devlin et~al.(2017)Devlin, Uesato, Bhupatiraju, Singh, rahman Mohamed,
  and Kohli]{devlin17a}
Jacob Devlin, Jonathan Uesato, Surya Bhupatiraju, Rishabh Singh, Abdel rahman
  Mohamed, and Pushmeet Kohli.
\newblock {R}obust{F}ill: Neural program learning under noisy {I}/{O}.
\newblock In \emph{International Conference on Machine Learning (ICML)},
  volume~70 of \emph{Proceedings of Machine Learning Research}, pp.\  990--998,
  2017.
\newblock URL \url{http://proceedings.mlr.press/v70/devlin17a.html}.

\bibitem[Elhage(2017)]{PBTIF}
Nelson Elhage.
\newblock Property-based testing is fuzzing, 2017.
\newblock URL \url{https://blog.nelhage.com/post/property-testing-is-fuzzing/}.

\bibitem[Ellis et~al.(2018{\natexlab{a}})Ellis, Morales, Sabl\'{e}-Meyer,
  Solar-Lezama, and Tenenbaum]{ECSQUARED}
Kevin Ellis, Lucas Morales, Mathias Sabl\'{e}-Meyer, Armando Solar-Lezama, and
  Josh Tenenbaum.
\newblock Learning libraries of subroutines for neurally\textendash guided
  bayesian program induction.
\newblock In S.~Bengio, H.~Wallach, H.~Larochelle, K.~Grauman, N.~Cesa-Bianchi,
  and R.~Garnett (eds.), \emph{Advances in Neural Information Processing
  Systems 31}, pp.\  7805--7815. Curran Associates, Inc., 2018{\natexlab{a}}.

\bibitem[Ellis et~al.(2018{\natexlab{b}})Ellis, Morales, Sablé-Meyer,
  Solar-Lezama, and Tenenbaum]{ellis18library}
Kevin Ellis, Lucas Morales, Mathias Sablé-Meyer, Armando Solar-Lezama, and
  Joshua~B. Tenenbaum.
\newblock Library learning for neurally-guided bayesian program induction.
\newblock In \emph{{NeurIPS}}, 2018{\natexlab{b}}.

\bibitem[Ellis et~al.(2019{\natexlab{a}})Ellis, Nye, Pu, Sosa, Tenenbaum, and
  Solar-Lezama]{WRITEEXECUTEASSESS}
Kevin Ellis, Maxwell Nye, Yewen Pu, Felix Sosa, Josh Tenenbaum, and Armando
  Solar-Lezama.
\newblock Write, execute, assess: Program synthesis with a repl.
\newblock \emph{arXiv preprint arXiv:1906.04604}, 2019{\natexlab{a}}.

\bibitem[Ellis et~al.(2019{\natexlab{b}})Ellis, Nye, Pu, Sosa, Tenenbaum, and
  Solar-Lezama]{ellis19}
Kevin Ellis, Maxwell Nye, Yewen Pu, Felix Sosa, Josh Tenenbaum, and Armando
  Solar-Lezama.
\newblock Write, execute, assess: Program synthesis with a {REPL}.
\newblock In \emph{{NeurIPS}}, 2019{\natexlab{b}}.

\bibitem[Feser et~al.(2015)Feser, Chaudhuri, and Dillig]{LAMBDASQUARED}
John~K Feser, Swarat Chaudhuri, and Isil Dillig.
\newblock Synthesizing data structure transformations from input-output
  examples.
\newblock In \emph{ACM SIGPLAN Notices}, volume~50, pp.\  229--239. ACM, 2015.

\bibitem[Freeman(1994)]{REFINEMENTTYPES}
Tim Freeman.
\newblock Refinement types ml.
\newblock Technical report, CARNEGIE-MELLON UNIV PITTSBURGH PA DEPT OF COMPUTER
  SCIENCE, 1994.

\bibitem[Gallant(2018)]{QCRUST}
Andrew Gallant.
\newblock Quickcheck for rust, 2018.
\newblock URL \url{https://github.com/BurntSushi/quickcheck}.

\bibitem[Gottschlich et~al.(2018)Gottschlich, Solar-Lezama, Tatbul, Carbin,
  Rinard, Barzilay, Amarasinghe, Tenenbaum, and Mattson]{THREEPILLARS}
Justin Gottschlich, Armando Solar-Lezama, Nesime Tatbul, Michael Carbin, Martin
  Rinard, Regina Barzilay, Saman Amarasinghe, Joshua~B Tenenbaum, and Tim
  Mattson.
\newblock The three pillars of machine programming.
\newblock In \emph{Proceedings of the 2nd ACM SIGPLAN International Workshop on
  Machine Learning and Programming Languages}, pp.\  69--80. ACM, 2018.

\bibitem[Gulwani(2011)]{FLASHFILL}
Sumit Gulwani.
\newblock Automating string processing in spreadsheets using input-output
  examples.
\newblock In \emph{Proceedings of the 38th Annual ACM SIGPLAN-SIGACT Symposium
  on Principles of Programming Languages}, POPL '11, pp.\  317--330, New York,
  NY, USA, 2011. ACM.
\newblock ISBN 978-1-4503-0490-0.
\newblock \doi{10.1145/1926385.1926423}.
\newblock URL \url{http://doi.acm.org/10.1145/1926385.1926423}.

\bibitem[Gulwani et~al.(2017)Gulwani, Polozov, Singh, et~al.]{RISHABHSURVEY}
Sumit Gulwani, Oleksandr Polozov, Rishabh Singh, et~al.
\newblock Program synthesis.
\newblock \emph{Foundations and Trends{\textregistered} in Programming
  Languages}, 4\penalty0 (1-2):\penalty0 1--119, 2017.

\bibitem[Heo et~al.(2019)Heo, Raghothaman, Si, and Naik]{heo19continuously}
Kihong Heo, Mukund Raghothaman, Xujie Si, and Mayur Naik.
\newblock Continuously reasoning about programs using differential bayesian
  inference.
\newblock In \emph{Programming Language Design and Implementation (PLDI)},
  2019.

\bibitem[Hochreiter \& Schmidhuber(1997)Hochreiter and Schmidhuber]{LSTM}
Sepp Hochreiter and J{\"u}rgen Schmidhuber.
\newblock Long short-term memory.
\newblock \emph{Neural computation}, 9\penalty0 (8):\penalty0 1735--1780, 1997.

\bibitem[Holser(2018)]{QCJAVA}
Paul Holser.
\newblock junit-quickcheck, 2018.
\newblock URL \url{https://github.com/pholser/junit-quickcheck/}.

\bibitem[Howard(1980)]{CURRYHOWARD}
William~A Howard.
\newblock The formulae-as-types notion of construction.
\newblock \emph{To HB Curry: essays on combinatory logic, lambda calculus and
  formalism}, 44:\penalty0 479--490, 1980.

\bibitem[Hypothesis(2018)]{QCPYTHON}
Hypothesis.
\newblock Hypothesis, 2018.
\newblock URL \url{https://github.com/HypothesisWorks/hypothesis}.

\bibitem[Irving et~al.(2016)Irving, Szegedy, Alemi, E{\'e}n, Chollet, and
  Urban]{DEEPMATH}
Geoffrey Irving, Christian Szegedy, Alexander~A Alemi, Niklas E{\'e}n,
  Fran{\c{c}}ois Chollet, and Josef Urban.
\newblock Deepmath-deep sequence models for premise selection.
\newblock In \emph{Advances in Neural Information Processing Systems}, pp.\
  2235--2243, 2016.

\bibitem[Kalyan et~al.(2018)Kalyan, Mohta, Polozov, Batra, Jain, and
  Gulwani]{NGDS}
Ashwin Kalyan, Abhishek Mohta, Oleksandr Polozov, Dhruv Batra, Prateek Jain,
  and Sumit Gulwani.
\newblock Neural-guided deductive search for real-time program synthesis from
  examples.
\newblock \emph{arXiv preprint arXiv:1804.01186}, 2018.

\bibitem[LeCun et~al.(1989)LeCun, Boser, Denker, Henderson, Howard, Hubbard,
  and Jackel]{CONVNETS}
Yann LeCun, Bernhard Boser, John~S Denker, Donnie Henderson, Richard~E Howard,
  Wayne Hubbard, and Lawrence~D Jackel.
\newblock Backpropagation applied to handwritten zip code recognition.
\newblock \emph{Neural computation}, 1\penalty0 (4):\penalty0 541--551, 1989.

\bibitem[Liang et~al.(2010)Liang, Jordan, and Klein]{HIERARCHICALBAYES}
Percy Liang, Michael~I Jordan, and Dan Klein.
\newblock Learning programs: A hierarchical bayesian approach.
\newblock In \emph{Proceedings of the 27th International Conference on Machine
  Learning (ICML-10)}, pp.\  639--646, 2010.

\bibitem[Luu(2015)]{AFLPLUSQUICKCHECK}
Dan Luu.
\newblock Afl + quickcheck = ?, 2015.
\newblock URL \url{https://danluu.com/testing/}.

\bibitem[MacIver(2017)]{WIPBT}
David~R. MacIver.
\newblock What is property based testing, 2017.
\newblock URL
  \url{https://hypothesis.works/articles/what-is-property-based-testing/}.

\bibitem[Manna \& Waldinger(1975)Manna and Waldinger]{KNOWLEDGEANDREASONING}
Zohar Manna and Richard Waldinger.
\newblock Knowledge and reasoning in program synthesis.
\newblock \emph{Artificial intelligence}, 6\penalty0 (2):\penalty0 175--208,
  1975.

\bibitem[Manna \& Waldinger(1971)Manna and Waldinger]{TOWARDAUTOMATIC}
Zohar Manna and Richard~J Waldinger.
\newblock Toward automatic program synthesis.
\newblock \emph{Communications of the ACM}, 14\penalty0 (3):\penalty0 151--165,
  1971.

\bibitem[Menon et~al.(2013)Menon, Tamuz, Gulwani, Lampson, and Kalai]{MLPBE}
Aditya Menon, Omer Tamuz, Sumit Gulwani, Butler Lampson, and Adam Kalai.
\newblock A machine learning framework for programming by example.
\newblock In \emph{International Conference on Machine Learning}, pp.\
  187--195, 2013.

\bibitem[Nye et~al.(2019)Nye, Hewitt, Tenenbaum, and
  Solar{-}Lezama]{INFERSKETCHES}
Maxwell~I. Nye, Luke~B. Hewitt, Joshua~B. Tenenbaum, and Armando
  Solar{-}Lezama.
\newblock Learning to infer program sketches.
\newblock In Kamalika Chaudhuri and Ruslan Salakhutdinov (eds.),
  \emph{Proceedings of the 36th International Conference on Machine Learning,
  {ICML} 2019, 9-15 June 2019, Long Beach, California, {USA}}, volume~97 of
  \emph{Proceedings of Machine Learning Research}, pp.\  4861--4870. {PMLR},
  2019.
\newblock URL \url{http://proceedings.mlr.press/v97/nye19a.html}.

\bibitem[Pierce \& Benjamin(2002)Pierce and Benjamin]{TAPL}
Benjamin~C Pierce and C~Benjamin.
\newblock \emph{Types and programming languages}.
\newblock 2002.

\bibitem[Pnueli \& Rosner(1989)Pnueli and Rosner]{REACTIVE}
Amir Pnueli and Roni Rosner.
\newblock On the synthesis of a reactive module.
\newblock In \emph{Proceedings of the 16th ACM SIGPLAN-SIGACT symposium on
  Principles of programming languages}, pp.\  179--190. ACM, 1989.

\bibitem[Polikarpova et~al.(2016)Polikarpova, Kuraj, and Solar-Lezama]{SYNQUID}
Nadia Polikarpova, Ivan Kuraj, and Armando Solar-Lezama.
\newblock Program synthesis from polymorphic refinement types.
\newblock In \emph{ACM SIGPLAN Notices}, volume~51, pp.\  522--538. ACM, 2016.

\bibitem[Shaw()]{LISP2}
D~Shaw.
\newblock Inferring lisp programs from examples.

\bibitem[Shin et~al.(2018)Shin, Kant, Gupta, Bender, Trabucco, Singh, and
  Song]{SHIFT}
Richard Shin, Neel Kant, Kavi Gupta, Chris Bender, Brandon Trabucco, Rishabh
  Singh, and Dawn Song.
\newblock Synthetic datasets for neural program synthesis.
\newblock 2018.

\bibitem[Solar-Lezama(2018)]{ARMANDOCOURSE}
Armando Solar-Lezama.
\newblock Introduction to program synthesis.
\newblock \url{https://people.csail.mit.edu/asolar/SynthesisCourse/TOC.htma},
  2018.
\newblock Accessed: 2018-09-17.

\bibitem[Spade \& Panaccio(2019)Spade and Panaccio]{OCCAM}
Paul~Vincent Spade and Claude Panaccio.
\newblock William of ockham.
\newblock In Edward~N. Zalta (ed.), \emph{The Stanford Encyclopedia of
  Philosophy}. Metaphysics Research Lab, Stanford University, spring 2019
  edition, 2019.

\bibitem[Summers(1977)]{LISP}
Phillip~D Summers.
\newblock A methodology for lisp program construction from examples.
\newblock \emph{Journal of the ACM (JACM)}, 24\penalty0 (1):\penalty0 161--175,
  1977.

\bibitem[Waldinger et~al.(1969)Waldinger, Lee, and International]{PROW}
R.J. Waldinger, R.C.T. Lee, and SRI International.
\newblock \emph{PROW: A Step Toward Automatic Program Writing}.
\newblock SRI International, 1969.
\newblock URL \url{https://books.google.com/books?id=3BITSQAACAAJ}.

\bibitem[Wang et~al.(2017)Wang, Cheung, and Bodik]{SQL}
Chenglong Wang, Alvin Cheung, and Rastislav Bodik.
\newblock Synthesizing highly expressive sql queries from input-output
  examples.
\newblock In \emph{Proceedings of the 38th ACM SIGPLAN Conference on
  Programming Language Design and Implementation}, PLDI 2017, pp.\  452--466,
  New York, NY, USA, 2017. ACM.
\newblock ISBN 978-1-4503-4988-8.
\newblock \doi{10.1145/3062341.3062365}.
\newblock URL \url{http://doi.acm.org/10.1145/3062341.3062365}.

\bibitem[Winston(1970)]{PBE}
Patrick~H. Winston.
\newblock Learning structural descriptions from examples.
\newblock Technical report, Cambridge, MA, USA, 1970.

\bibitem[Zohar \& Wolf(2018)Zohar and Wolf]{GARBAGECOLLECTOR}
Amit Zohar and Lior Wolf.
\newblock Automatic program synthesis of long programs with a learned garbage
  collector.
\newblock In S.~Bengio, H.~Wallach, H.~Larochelle, K.~Grauman, N.~Cesa-Bianchi,
  and R.~Garnett (eds.), \emph{Advances in Neural Information Processing
  Systems 31}, pp.\  2094--2103. Curran Associates, Inc., 2018.

\end{thebibliography}
\bibliographystyle{iclr2020_conference}

\newpage
\newpage

\appendix
\section{Appendix}
\subsection{Further Details on the Baseline Synthesizer}
\label{app:baseline}
This section contains details on the baseline synthesizer that did not fit into
the main text.
Figure \ref{fig:algorithm} gives a more formal description of the basic
synthesis algorithm.
Listing \ref{listing:trajectory} shows an example trajectory of partial program
expansions.

\begin{figure}[tb]
  \centering
       \begin{algorithm}[H]
        \SetAlgoLined
        \SetKwFunction{GetLowestCostPartial}{GetLowestCostPartial}
        \SetKwFunction{HasHoles}{HasHoles}
        \SetKwFunction{ExpandOneHole}{ExpandOneHole}
        \SetKwFunction{TestAgainstSpec}{TestAgainstSpec}
        \KwData{A PBE spec and a synthesizer configuration}
        \KwResult{A program satisfying the specification (hopefully!)}
        \texttt{Queue.push}(\texttt{hole} $:: \tau_{in} \rightarrow \tau_{out}$)\;
         \While{\texttt{Queue} is not empty}{
          $\texttt{partial\_program} \leftarrow$ \GetLowestCostPartial{$\texttt{Queue}$} \;
          \If{\HasHoles{$\texttt{partial\_program}$}}{
           \ExpandOneHole{$\texttt{partial\_program}$}\;
           }
          \Else{
           \TestAgainstSpec{$\texttt{partial\_program}$}\;
           }
         }
        \end{algorithm}
  \caption{
  The top-down synthesizer that we use as a baseline in this work.
  In a loop until a satisfying program is found or we run out of time, we pop
  the lowest-cost partial program from the queue of all partial programs, then
  we fill in the holes in all ways allowed by the type system, pushing each new
  partial program back onto the queue.
  If there are no holes to fill, the program is complete, and we check it
  against the spec.
  The cost of a partial program is the sum of the costs of
  its pool elements, plus a lower bound on the cost of filling each of its
  typed holes, plus the sum of the costs of a few special operations such as
  tuple construction and lambda abstraction.
  }
\label{fig:algorithm}
\end{figure}

\begin{lstlisting}[language=Haskell, caption={
The trajectory the synthesizer took to generate the swap function, which
just swaps the two elements of a tuple.
Since it knows it needs to take a tuple of ints as an argument and return
a tuple of ints, it starts with a hole of type $(\texttt{Int}, \texttt{Int})$
in line 1.
It then converts that hole into a tuple of holes, both of type $\texttt{Int}$
in line 2, fills one of the holes with a reference to one of the arguments in
line 3, and fills in the final hole with a reference to the other argument in
line 4.
Note that this listing doesn't show all programs attempted, it just shows
the sequence of partial programs that led to the final solution.
}, label=listing:trajectory]
$1 \:(Int, Int)->(Int, Int) (a2, a3) {?}
$2 \:(Int, Int)->(Int, Int) (a2, a3) {(?, ?)}
$2 \:(Int, Int)->(Int, Int) (a2, a3) {(a3, ?)}
$2 \:(Int, Int)->(Int, Int) (a2, a3) {(a3, a2)}
\end{lstlisting}

\end{document}